\documentclass[twocolumn,showpacs,preprintnumbers,amsmath,amssymb]{revtex4}
\usepackage{cases}
\usepackage{amssymb}
\usepackage{txfonts}
\usepackage{amsmath}

\usepackage{graphicx}
\usepackage{dcolumn}
\usepackage{bm}

\begin{document}

\preprint{}

\title{Zitterbewegung-like effect near the Dirac point in metamaterials and photonic crystals
}
\author{Xiaohui Ling}
\author{Zhixiang Tang}\email{tzx@hnu.edu.cn}
\author{Hailu Luo}
\author{Huimin Dong}
\author{Zhaoming Luo}
\author{Shuangchun Wen}\email{scwen@hnu.cn}
\author{Dianyuan Fan}
\affiliation{Key Laboratory for Micro/Nano Opto-Electronic Devices
of Ministry of Education, School of Information Science and
Technology, Hunan University, Changsha 410082, China}
\date{\today}

\begin{abstract}
We present a physical explanation of Zitterbewegung-like effect near
the zero-refractive-index point in a metamaterial slab in this
paper. Between the negative and positive refractive index regions
centered at the zero-refractive-index point, the transmittance
spectrum distribution of the metamaterial slab is asymmetrical. When
a symmetrical pulse propagates through the metamaterial slab, its
transmitted spectrum becomes asymmetrical due to the asymmetry of
the transmittance spectrum of the slab, leading to a transmitted
pulse with an asymmetrical temporal shape. The asymmetry manifests a
kind of temporally tailed oscillations, $\textit{i.e.}$, the
Zitterbewegung-like effect. Further, the effect of the temporal and
spatial widths of pulse, and the thickness of metamaterial slab on
the tailed oscillations of the transmitted pulse has also been
discussed. Our results agree well with what the other researchers
obtained on the strength of relativistic quantum concepts; however,
the viewpoint of our analysis is classical and irrelevant to
relativistic quantum mechanics.
\end{abstract}


\maketitle

\section{Introduction}\label{SecI}
The propagation of electromagnetic waves near the Dirac point (DP)
in photonic crystals and metamaterials has been investigated
intensively in recent
years~\cite{Haldane2008,Raghu2008,Sepkhanov2007,Diem2010,Sepkhanov2008,
Zhang2008PLA,Peleg2007,Zhang2008PRL1,Zhang2008PRL2,Wang2009OL,Wang2009EPL}.
With some special structure designs of both kinds of artificial
materials, two energy bands touch each other, forming a pair of
cones, namely, so-called ``Dirac point'' in optics. For a
monochromatic electromagnetic wave propagating near the DPs, some
abnormal transmission properties have been discovered, such as
pseudo-diffusive scaling
~\cite{Sepkhanov2007,Diem2010,Sepkhanov2008, Zhang2008PLA} and
conical diffraction~\cite{Peleg2007}. The propagation of optical
pulses with the central frequency at the DP in a photonic crystal or
metamaterial slab has also been investigated by some research groups
who show that temporally tailed oscillations of the transmitted
pulses should appear ~\cite{Zhang2008PRL1,Wang2009EPL}. In their
papers, this kind of oscillation is called Zitterbewegung (ZB)
effect for photon.

The existence of ZB effect was first proposed by Erwin
Schr\"{o}dinger for relativistic electrons in free space, and the
interference between the positive and negative energy states was
considered as the origin~\cite{Schrodinger1930}. ZB effects in
crystals, superconductors, semiconductor nanostructures, graphene,
and other physical systems have also been
demonstrated~\cite{Cannata1990,Cannata1991,Ferrari1990,Schliemann2005,Shen2005,Jiang2005,Brusheim2006,Zawadzki2005,Zawadzki2006,Katsnelson2006,
Lurie1970,Cserti2006,Vaishnav2008,Lamata2007,Wang2008}. And ZB
effect for photon has been proposed in two-dimensional photonic
crystal for the first time by simulation analysis by
Zhang~\cite{Zhang2008PRL1}, in which he used a plane wave with
Guassian shape in time domain, passing through an open slit, to
excite the photonic crystal slab. In another paper for the case of
sonic crystals, Zhang and his co-worker demonstrated the ZB effect
in experiment and mentioned that the origin of this effect was the
interference between two linear modes around the
DP~\cite{Zhang2008PRL2}. Shortly thereafter, DP in the
negative-zero-positive index metamaterial (NZPIM) was first proposed
by Wang $\textit{et al.}$~\cite{Wang2009OL}. Thereupon the ZB effect
of optical pulses in NZPIM was characterized~\cite{Wang2009EPL}.
Both the research groups named the temporally tailed oscillations of
the transmitted pulses as ZB effect and thus explained it on the
strength of some concepts of the relativistic quantum mechanics,
such as DP and Dirac equation. This is undoubtedly an interesting
and helpful understanding of the underlying mechanism of the effects
in photonic crystal and metamaterial. However, we think that it may
be not suitable to use ZB directly as the name of the tailed
oscillations of optical pulses. So, here, we introduce
Zitterbewegung-like (ZB-like) instead.

In this work, based on another point of view which is irrelevant to
the relativistic quantum mechanics, we present a physical
explanation of temporally tailed oscillations (or ZB-like effect) of
the pulses propagating near the zero-refractive-index point (called
DP in Refs.~\cite{Wang2009OL,Wang2009EPL}) in NZPIM using the
spatio-temporal filtering analysis. Due to the asymmetrical
transmittance spectrum distribution between both sides of the
zero-refractive-index point, a symmetrical pulse passed through a
NZPIM slab undergoes an asymmetrical filtering effect, which leads
to the asymmetrical temporal shape of the transmitted pulse, in
other words, the tailed oscillations. Further, the effect of the
temporal and spatial widths of pulse, and the thickness of the NZPIM
slab on the tailed oscillations of the transmitted pulse can be
characterized well. The analysis method we use is similar to that of
previous
works~\cite{Zhang2008PRL1,Zhang2008PRL2,Wang2009OL,Wang2009EPL} on
dealing with some detail problems, however, our viewpoint, based on
classical spectral filtering analysis that is not related to the
relativistic quantum mechanics, is totally different from theirs.
Even more interesting is the fact that the results we obtained are
in good agreement with those in Refs.~\cite{Zhang2008PRL1}
and~\cite{Wang2009EPL}. Using this analysis method, ZB-like effect
in two-dimensional photonic crystals could also be explained.
\section{Model and theory}\label{SecII}
We take the homogenous NZPIM slab as the analysis sample. This kind
of metamaterial has already been demonstrated by theoretical and
experimental researches based on liquid crystals and $\Omega$-shaped
microstructures from GHz to visible
regions~\cite{Khoo2006,Werner2007,FZhang2008,Bossard2008,Kwon2007}.
For simplicity, Drude model is chosen as the parameter of the
permittivity and the permeability of the NZPIM
slab~\cite{Wang2009OL,Ziolkowski2004}:
\begin{eqnarray}
\epsilon\left(\omega\right)=1-\frac{\omega_{pe}^2}{\omega^2+\textit{i}\gamma_e\omega},~
\mu\left(\omega\right)=1-\frac{\omega_{pm}^2}{\omega^2+\textit{i}\gamma_m\omega},\label{one}
\end{eqnarray}
where $\omega_{pe}$ and $\omega_{pm}$ are controllable electronic
and magnetic plasma frequencies, and $\gamma_e$ and $\gamma_m$ are
the damping rates. As shown in~\cite{FZhang2008}, with suitable
structure parameters, $\omega_{p}=\omega_{pe}=\omega_{pm}$ can be
achieved in a back-to-back $\Omega$-shaped metamaterial in GHz
region. When $\omega=\omega_{p}=\omega_{pe}=\omega_{pm}$ and
$\gamma=\gamma_e=\gamma_m\ll\omega_{p}$, $\epsilon(\omega)$ and
$\mu(\omega)$ are nearly equal to zero, simultaneously. Thus, the
effective refractive index of the sample varies from negative,
through zero, then to positive with the increase of frequency. In
such case, the so-called DP is the zero point of effective
refractive index.

\begin{figure}
\includegraphics[width=8.2cm]{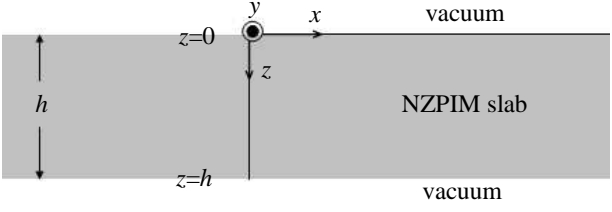}
\caption{\label{Fig1} Schematic of the NZPIM slab in vacuum with
thickness $\textit{h}$ along $\textit{z}$ direction. It is assumed
that the slab is infinite along $\textit{x}$ and $\textit{y}$
direction.}
\end{figure}

We consider a paraxial, coherent, TE-polarized, spatially and
temporally Guassian pulse injecting into the sample along
$\textit{z}$ direction in vacuum as shown in the Fig.~\ref{Fig1}. It
is assumed that the slab sample is infinite along $\textit{x}$ and
$\textit{y}$ direction. The refractive
index is $\textit{n}_1(\omega)=\sqrt{\epsilon(\omega)\mu(\omega)}
=1-\omega_p^2/(\omega^2+\textit{i}\gamma\omega)$ and the thickness
of the sample is $\textit{h}$. The function of the pulse at the
initial plane $\textit{z}=\textit{z}_0<0$ [while plane
$\textit{z}=0$ is the interface between vacuum ($\textit{n}_0=1$)
and the NZPIM slab sample] is given as
\begin{eqnarray}
\textit{E}_{\textit{i}}\left(\textit{x},\textit{t},\textit{z}_0\right)=\exp\left(-\frac{t^2}{\Gamma^2}-\frac{\textit
{x}^2}{\textit{W}_0^2}\right)\exp\left(-\textit{i}\omega_0\textit{t}\right).\label{two}
\end{eqnarray}
This pulse can be decomposed into its monochromatic Fourier
components and further be expanded in angular spectra of
monochromatic plane-wave components
\begin{eqnarray}
\textit{E}_{\textit{i}}\left(\textit{k}_\textit{x},\omega,\textit{z}_0\right)=\frac{\Gamma\textit{W}_0}{4\pi}
\exp\left[-\frac{\Gamma^2\left(\omega-\omega_0\right)^2}{4}\right]\exp\left(-\frac{\textit{k}_\textit{x}^2\textit{W}_0^2}{4}\right),\label{three}
\end{eqnarray}
where $\Gamma$ is the temporal half-width, $\textit{W}_0$ is the
spatial half-width,
$\textit{k}_\textit{x}=(\textit{n}_0\omega/\textit{c})$$\sin\theta$,
$\textit{c}$ is the light velocity in vacuum, and $\omega_0$ is the
central angular frequency of the pulse.

The transmission coefficient
$\textit{t}(\theta,\omega)=|\textit{t}(\theta,\omega)|\exp[\textit{i}\varphi(\theta,\omega)]$
of the sample can be deduced from the electromagnetic continuity
conditions of the two interfaces (plane $\textit{z}=0$ and
$\textit{z}=\textit{h}$ in Fig.~\ref{Fig1})~\cite{Born1999}. In this
paper, for the TE-polarized incident wave,
$\textit{t}(\theta,\omega)$ can be written as
\begin{eqnarray}
\textit{t}(\theta,\omega)=\frac{2}{2\cos\delta-\textit{i}\sin\delta\left(\frac{\cos\theta_1}{\cos\theta}+
\frac{\cos\theta}{\cos\theta_1}\right)}.\label{four}
\end{eqnarray}
Here,
$\theta_1=\arcsin[\textit{n}_0\sin\theta/\textit{n}_1(\omega)]$ is
the refractive angle of the angular spectrum component in the sample
which can be obtained from Snell's law and
$\delta=\omega\textit{n}_1(\omega)\textit{h}\cos\theta_1/\textit{c}$
stands for the phase change when electromagnetic wave cross the
sample~\cite{Born1999}.
\begin{figure}
\includegraphics[height=14cm]{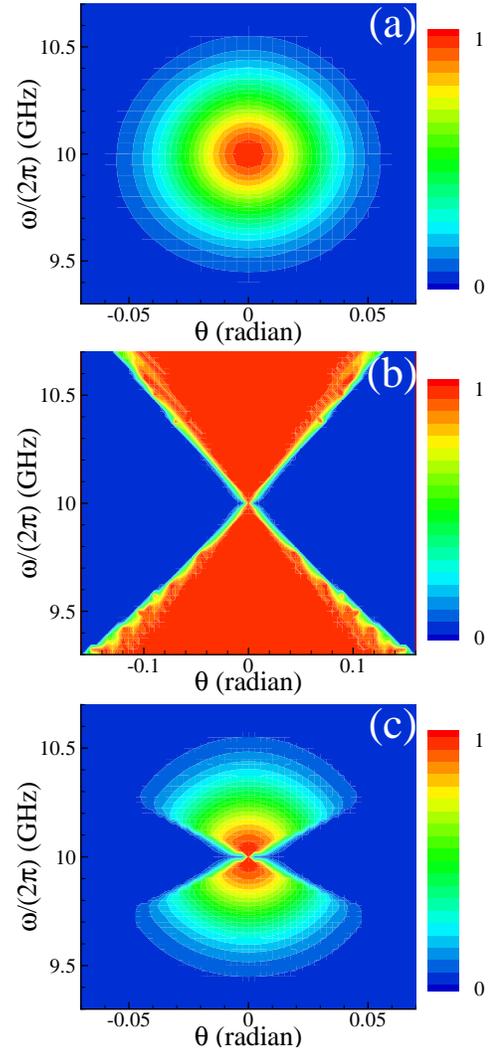}
\caption{\label{Fig2}(Color online) (a) The spatio-temporal spectrum
of incident Guassian pulse with $\Gamma=1.0$ns,
$\textit{W}_0=10\lambda_0$ ($\lambda_0=0.03$m), and $\textit{f}_0=
10$GHz. (b) Distribution of the spectral transmittance
($|\textit{t}(\theta,\omega)|$) for a sample with
$\textit{h}=40\lambda_0$ and $\gamma=\gamma_e=\gamma_m=10^4$Hz, the
blue part denotes the stop-band and the rest denotes the pass-band.
When $\omega=\omega_p=2\pi\times10$GHz, the refractive index is
zero. (c) The spatio-temporal spectrum of the transmitted pulse.}
\end{figure}

In this case, the transmission coefficient is equivalent to the
frequency response of a linear
filter~\cite{Desbois1973,Froehly1983,Weiner1995,Weiner2000,Vallius2002}.
Therefore the spectral function at the exit end can be written as
$\textit{E}_\textit{t}(\theta,\omega,\textit{z})=\textit{E}_\textit{i}(\theta,\omega,0)
\textit{t}(\theta,\omega)$.
$\textit{E}_\textit{i}(\theta,\omega,0)=\textit{E}_\textit{i}(\theta,\omega,\textit{z}_0)\exp(-\textit{i}\textit{k}_\textit{z}\textit{z}_0)$
is the function of spatio-temporal spectrum of the pulse arriving at
the plane $\textit{z}=0$, where
$\textit{k}_\textit{z}=\textit{k}_0\cos\theta$
($\textit{k}_0=\textit{n}_0\omega/\textit{c}$ is the wave vector in
vacuum). We take the spatial and temporal half-width of the incident
pulse as 10$\lambda_0$ ($\lambda_0=\textit{c}/\textit{f}_0$ is
central wavelength of the pulse in vacuum) and 1ns, respectively,
with the central frequency $\textit{f}_0$ (=10GHz) for an example.
The spatio-temporal spectrum of the incident pulse and the
transmittance spectrum distribution $|\textit{t}(\theta,\omega)|$
with slab thickness $\textit{h}=40\lambda_0$ are plotted in
Figs.~\ref{Fig2}(a) and \ref{Fig2}(b), respectively. Unambiguously,
the output spectrum is tailored by the transmittance spectrum as
shown in Fig.~\ref{Fig2}(c). Some spatio-temporal spectrum
components are filtered by the stop-band of the transmittance
spectrum. In order to obtain the spatio-temporal shape
$\textit{E}_\textit{t}(\textit{x},\textit{t},\textit{z})$ of the
transmitted pulse, the inverse Fourier transform of
$\textit{E}_\textit{t}(\theta,\omega,\textit{z})$ has been
calculated. But it is difficult to solve such a complicated
expression analytically, so numerical integration method is used
here. In our analysis, we only need to consider the temporal shape
of the transmitted pulses [we consider
$\textit{E}_\textit{t}(\textit{x}=0$, $\textit{t},\textit{z})$]
because ZB-like effect is the temporally tailed oscillations.

\section{Results and discussions}\label{SecIII}
As is shown in Fig.~\ref{Fig2}(b), the pass-band of the
transmittance spectrum gradually shrinks to the
zero-refractive-index point from both sides of it, forming a
double-cone band. The underlying mechanism is that toward the
zero-refractive-index point from both sides, the critical angle of
the total reflection at the interface $\textit{z}=0$ decreases,
thereby only normal incident angular spectra of the pulse can pass
through the sample at the zero-refractive-index point, which can be
deduced from Snell's law. Even more importantly, the transmittance
spectrum is asymmetrical between negative and positive refractive
index regions in respect that
$\textit{n}_1(\omega)=1-\omega_p^2/(\omega^2+\textit{i}\gamma\omega)$
is asymmetrical for the two regions. The asymmetrical transmittance
spectrum gives an asymmetrical filtering effect to the incident
spatio-temporal spectrum. This asymmetry in frequency domain is the
origin of the tailed oscillations in time domain. In a sense, a
pulse shape with tailed oscillations could be seen as a kind of
asymmetry. At the same time, the transmittance spectrum is
symmetrical for negative and positive incident angle $\theta$,
deduced from Eq.~(\ref{four}). Thus, the spatial shape of the pulse
is symmetrical (seen Fig.~\ref{Fig2} in Ref.~\cite{Wang2009EPL}). In
addition, it is worth noting that the temporally tailed oscillations
will also arise when the central frequency of the pulse diverge from
the zero-refractive-index point slightly, on condition that the
asymmetrically filtering effect still holds.

\begin{figure}
\includegraphics[height=14cm]{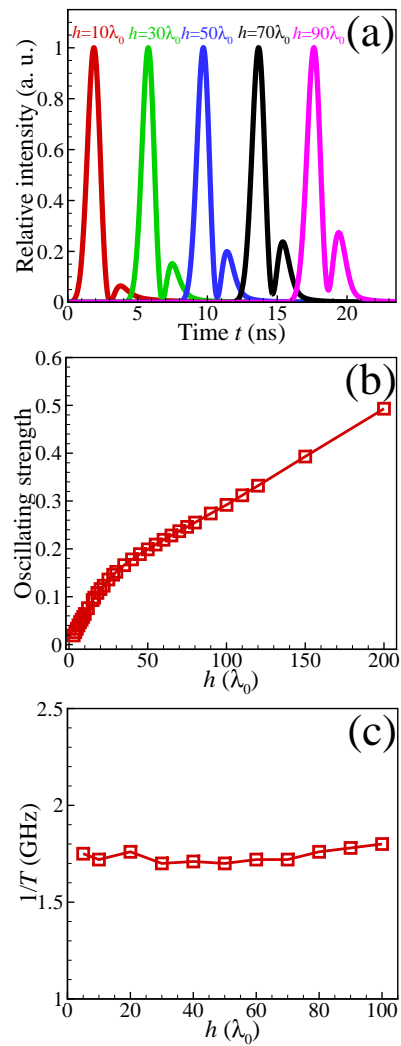}
\caption{\label{Fig3}(Color online) (a) Relative intensity of the
transmitted pulse from the calculating data of the inverse Fourier
transform with sample thickness 10$\lambda_0$, 30$\lambda_0$,
50$\lambda_0$, 70$\lambda_0$, and 90$\lambda_0$, respectively, with
the fixed $\Gamma=1.0$ns and $\textit{W}_0=10\lambda_0$. (b)
Oscillating strength versus $\textit{h}$. (c) Oscillating frequency
1/$\textit{T}$ versus $\textit{h}$.}
\end{figure}

Equation~(\ref{four}) determines the sample's transmittance spectrum
that is a function of the sample thickness $\textit{h}$. As
$\textit{h}$ increases, the transition region from the pass-band to
stop-band becomes steeper and steeper with increasingly severe
oscillations, which will influence the filtering effect and, then,
the oscillating properties of the transmitted pulse. In
Fig.~\ref{Fig3}(a), we plot the relative intensity
$|(\textit{E}_t(\textit{x}=0,\textit{t},\textit{z})/\textit{E}_t(\textit{x}=0,\textit{t},\textit{z})_{max}|^2$
of the transmitted pulses with the sample thickness
$\textit{h}=10\lambda_0$, 30$\lambda_0$, 50$\lambda_0$,
70$\lambda_0$, and 90$\lambda_0$, respectively, while $\Gamma=1.0$ns
and $\textit{W}_0=10\lambda_0$ are fixed. The initial center of the
pulse is assumed at $\textit{z}_0=-20\lambda_0$. The oscillating
strength (the ratios of relative intensities between the second peak
and the first peak) of the tailed oscillations increases with
$\textit{h}$ while the oscillating period $\textit{T}$ (time spacing
between the second peak and the first peak) remains nearly
unchanged, which can be seen from Figs.~\ref{Fig3}(b) and
\ref{Fig3}(c). When $\textit{h}$ increases to several hundreds of
$\lambda_0$, the dispersion effect is not negligible and should
seriously distort the temporal shape of the pulse.

\begin{figure}
\includegraphics[height=14cm]{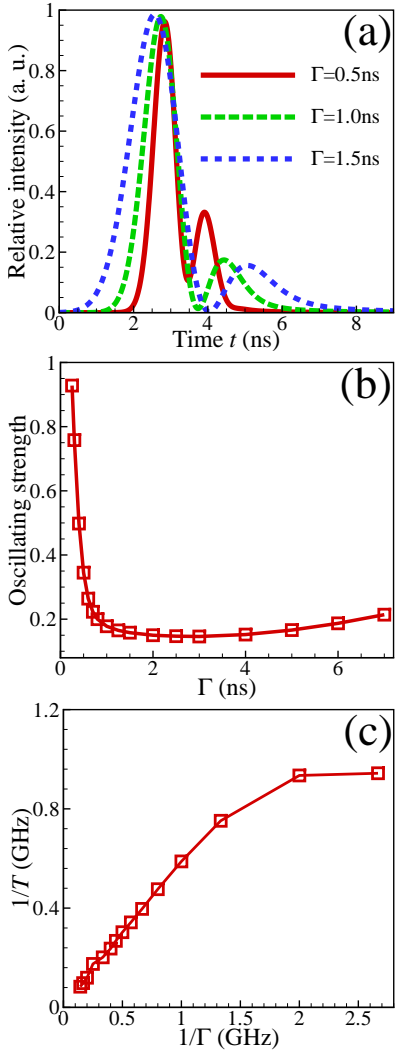}
\caption{\label{Fig4}(Color online) (a) Relative intensity of the
transmitted pulse with different temporal half-width 0.5ns, 1.0ns,
and 1.5ns, respectively, with $\textit{W}_0=10\lambda_0$ and
$\textit{h}=40\lambda_0$. (b) Oscillating strength versus $\Gamma$.
(c) Oscillating frequency 1/$\textit{T}$ versus 1/$\Gamma$.}
\end{figure}

Pulse with different temporal half-width $\Gamma$ will undergo
different responses when propagating in the NZPIM slab sample. The
incident spatio-temporal spectrum varies with $\Gamma$, thus results
in different filtering effects. Near the zero-refractive-index
point, the dispersion is considered to be approximately
linear~\cite{Wang2009OL}, however, the linear dispersion
approximation is not valid when $\Gamma$ becomes very small (few
optical cycles in time domain but extremely large bandwidth in
frequency domain). In Fig.~\ref{Fig4}(a), we plot the relative
intensity of the transmitted pulses with $\Gamma=0.5$ns, 1.0ns, and
1.5ns, respectively, while $\textit{W}_0=10\lambda_0$ and
$\textit{h}=40\lambda_0$ are fixed. Figure~\ref{Fig4}(b) illustrates
the relationship between the oscillating strength and $\Gamma$. As
$\Gamma<0.8$ns (only several optical cycles), the oscillating
strength varies sharply due to the influence of high-order
dispersions. While for $\Gamma>0.8$ns, the high-order dispersions
are negligible, so the oscillating strength has a relatively flat
dependence on $\Gamma$ in the framework of filtering effect alone.
The oscillating frequency 1/$\textit{T}$ increases linearly with the
1/$\Gamma$ under the linear dispersion approximation as shown in
Fig.~\ref{Fig4}(c) (when $1/\Gamma>1.25$GHz, the dispersion is no
longer linear).

\begin{figure}
\includegraphics[height=14cm]{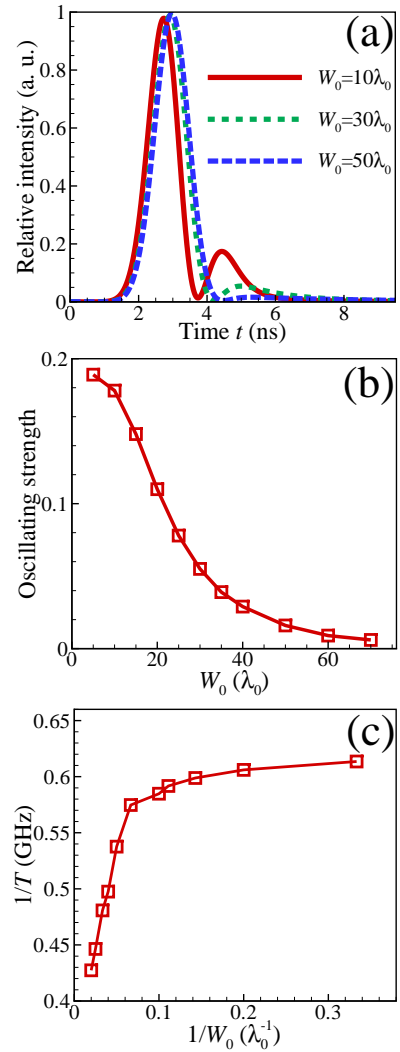}
\caption{\label{Fig5}(Color online) (a) Relative intensity of the
transmitted pulse with different spatial half-width 10$\lambda_0$,
20$\lambda_0$, and 30$\lambda_0$, respectively, with $\Gamma=1$ns
and $\textit{h}=40\lambda_0$. (b) Oscillating strength versus
$\textit{W}_0$. (c) Oscillating frequency 1/$\textit{T}$ versus
1/$\textit{W}_0$.}
\end{figure}

Though the ZB-like effect occurs in time domain, the spatial width
of the pulse does, indeed, influence the oscillating properties.
When $\textit{W}_0$ increases, the spatial part of the incident
spatio-temporal spectrum becomes narrower and narrower
($\theta$-axis direction in Fig.~\ref{Fig2}(a)). If we consider the
limiting situation, that is, $\textit{W}_0$ tends to infinity, the
pulse only has the angular spectra in normal incident direction,
under the circumstances, no spectrum component should be filtered by
the transmittance spectrum. So the temporally tailed oscillations
would be too weak (nearly vanished) to observe as long as
$\textit{W}_0$ becomes enough large. In Fig.~\ref{Fig5}(a), we plot
the relative intensity of the transmitted pulses with
$\textit{W}_0=10\lambda_0$, 30$\lambda_0$, and 50$\lambda_0$,
respectively, while $\Gamma=1$ns and $\textit{h}=40\lambda_0$ are
fixed. Seen from Fig.~\ref{Fig5}(b), we know the oscillating
strength decreases with the $\textit{W}_0$. The oscillating
frequency 1/$\textit{T}$ raises sharply with small values of
1/$\textit{W}_0$ and more flatly after coming up to a certain value
(about 0.06, see Fig.~\ref{Fig5}(c)).

It should be pointed out that, although we have only discussed the
optical ZB-like effect in metamaterial in the above analysis, the
similar phenomena in two-dimensional photonic crystals could be
interpreted in the same way. In a recent
publication~\cite{Diem2010}, an extremely similar transmittance
spectrum in two-dimensional hexagonal photonic crystals near the DP
has been revealed.
\section{Conclusions}
In conclusion, based on the spatio-temporal filtering analysis, we
have presented a physical explanation for the ZB-like effect in
NZPIM. When a symmetrical pulse passes through a NZPIM slab near the
zero-refractive-index point, its spatio-temporal spectrum will be
filtered asymmetrically for negative and positive refractive index
regions, which results in the ZB-like effect in time domain. The
strength of tailed oscillations depends on the sample thickness
$\textit{h}$, temporal half-width $\Gamma$, and the spatial
half-width $\textit{W}_0$ while the oscillating frequency only
relates to $\Gamma$ and $\textit{W}_0$, and do not change with
$\textit{h}$. The results we have got are very similar to what the
authors of Refs.~\cite{Zhang2008PRL1} and~\cite{Wang2009EPL}
obtained based on the concepts of relativistic quantum mechanics,
however, our analysis is a classical method which is irrelevant to
the relativistic quantum mechanics.
\begin{acknowledgements}
The authors are sincerely grateful to the anonymous referees whose
valuable suggestions allowed us to substantially improve the quality
of exposition. This work was supported by the Hunan Provincial
Natural Science Foundation of China (No. 08JJ3121) and the
Specialized Research Fund for the Doctoral Program of Higher
Education of China (No. 20090161120029).
\end{acknowledgements}

\end{document}